\documentclass[aps,preprint,showpacs,showkeys]{revtex4}
\usepackage{amsfonts}
\usepackage{amsmath}
\usepackage{amssymb}
\usepackage{graphicx}

\begin{document}

\title{Relating pseudospin and spin symmetries through chiral transformation with tensor interaction}\author{L.B. Castro}\email[ ]{luis.castro@pgfsc.ufsc.br}
\affiliation{Departamento de F\'{\i}sica CFM,
Universidade Federal de Santa Catarina, CP. 476, CEP 88.040-900, Florian\'{o}polis, Santa Catarina, Brazil} 

\begin{abstract}

We address the behavior of the Dirac equation with scalar ($S$), vector ($V$) and tensor ($U$) interactions under the $\gamma^{5}$ discrete chiral transformation. Using this transformation we can obtain from a simple way solutions for the Dirac equation with spin ($\Delta=V-S=0$) and pseudospin ($\Sigma=V+S=0$) symmetries including a tensor interaction. As an application, the Dirac equation with scalar, vector and tensor Cornell radial potentials is considered and the correct solution to this problem is obtained.

\end{abstract}

\maketitle

%=====================================================
\section{Introduction}

The pseudospin symmetry has been introduced in nuclear physics many years ago \cite{ari}-\cite{hecht}, in order to account for the degeneracies of orbitals in single-particle spectra. It is known that the spin and pseudospin symmetries correspond to SU($2$) symmetries of a Dirac Hamiltonian with vector ($V$) and scalar ($S$) potentials. Also, it is known that the spin symmetry occurs in the spectrum of a meson with one heavy quark \cite{gi1} and anti-nucleon bound in a nucleus \cite{gi2}, and the pseudospin symmetry occurs in the spectrum of nuclei \cite{gi3},\cite{gi5}. In \cite{zhou}-\cite{castro2} was reported that pseudospin and spin symmetries are connected by charge conjugation. This was shown explicitly for harmonic oscillator potentials in (1+1) dimensions \cite{castro1}. Also, it is shown in \cite{castro1} a connection of pseudospin and chiral symmetries in (1+1) dimensions. In recent years, some authors have extended the research field for pseudospin and spin symmetries including a tensor interaction. The tensor interaction has been used in studies of nuclear properties with effective Lagrangians including Relativistic Mean-Field (RMF) theories \cite{furn} and in the relativistic Hartree approach model \cite{mao}. Those works suggests that the tensor interaction could have a significant contribution to pseudospin splittings in nuclei. In \cite{chia} shows that the tensor interaction can change strongly the spin-orbit term. The connection of pseudospin and spin symmetries by charge conjugation including a tensor interaction has also been studied in \cite{castro3} and \cite{castro4}. However, a clear connection between pseudospin and spin symmetries obtained by a discrete chiral transformation including a tensor interaction has not been established. Therefore, we believe that this connection deserves to be explored.

The main motivation of this paper is inspired on the results obtained in Ref. \cite{castro1}. As a natural extension we address the behavior of the Dirac equation with scalar ($S$), vector ($V$) and tensor ($U$) interactions under the $\gamma^{5}$ discrete chiral transformation. Using this transformation we can obtain from a simple way solutions for the Dirac equation with spin ($\Delta=V-S=0$) and pseudospin ($\Sigma=V+S=0$) symmetries including a tensor interaction. As an application, the Dirac equation with scalar, vector and tensor Cornell radial potentials is considered. The radial equation for this problem is mapped into a Schr\"{o}dinger-like equation embedded in a three-dimensional harmonic oscillator plus a Cornell potential. We use this opportunity to present the correct solution to this problem in a more transparent way.

\section{Dirac equation with scalar, vector and tensor interactions}

The time-independent Dirac equation for a fermion with scalar ($S$), vector ($V$) and tensor ($U$) interactions is given by ($\hbar =c=1$)%
\begin{equation}\label{deh}
    H\Psi=E\psi
\end{equation}%
\noindent where
\begin{equation}\label{ham}
   H= \vec{\alpha}\cdot\vec{p}+\beta\left( m+S(r) \right)+V(r)-i\beta\vec{\alpha}\cdot\hat{r}\,U(r)\,.
\end{equation}

\noindent Using the combinations $\Sigma=V+S$ and $\Delta=V-S$, we can rewrite the Hamiltonian (\ref{ham}) as
\begin{equation}\label{ham2}
    H=\vec{\alpha}\cdot\vec{p}+\beta m + \frac{I+\beta}{2}\,\Sigma+\frac{I-\beta}{2}\,\Delta-i\beta\vec{\alpha}\cdot\hat{r}\,U(r)
\end{equation}

\subsection{Chiral transformation}

The chiral operator is the matrix $\gamma^{5}=i\gamma^{0}\gamma^{1}\gamma^{2}\gamma^{3}$ and therefore under the discrete chiral transformation the spinor is transformed as $\Psi_{\chi}=\gamma^{5}\Psi$ and the transformed Hamiltonian $H_{\chi}=\gamma^{5}H\gamma^{5}$ is
\begin{equation}\label{hamchi}
    H_{\chi}=\vec{\alpha}\cdot\vec{p}-\beta\left( m+S(r) \right)+V(r)+i\beta\vec{\alpha}\cdot\hat{r}\,U(r)
\end{equation}

\noindent we can see that the discrete chiral transformation changes the sign of the mass and of the scalar and tensor potentials because $\gamma^{5}$ commutes with $\vec{\alpha}$ and anticommutes with $\beta$. In term of the combinations $\Sigma$ and $\Delta$, this means that $\Sigma$ turns into $\Delta$ and vice-versa.

\subsection{Equation of motion}

Now, we follow the same procedure of Ref. \cite{castro4} and use the projectors $P_{\pm}=\left( I\pm\beta \right)/2$. Applying $P_{\pm}$ to the left of the Dirac equation (\ref{deh}) and defining $\Psi_{\pm}=P_{\pm}\Psi$ we obtain
\begin{equation}\label{eqd2}
    \vec{\alpha}\cdot\vec{p}\,\Psi_{\mp}+\left[ V(r)\pm\left( m+S(r) \right) \right]\Psi_{\pm}\mp i\vec{\alpha}\cdot\hat{r}\,U(r)\Psi_{\mp}=E\Psi_{\pm}
\end{equation}

\noindent or
\begin{eqnarray}\label{eqd3}
  \left( \vec{\alpha}\cdot\vec{p}-i\vec{\alpha}\cdot\hat{r}\,U(r) \right)\Psi_{-} &=& \left( E-m-\Sigma \right)\Psi_{+}\label{eqd3a} \\
  \left( \vec{\alpha}\cdot\vec{p}+i\vec{\alpha}\cdot\hat{r}\,U(r) \right)\Psi_{+} &=& \left( E+m-\Delta \right)\Psi_{-}\label{eqd3b}
\end{eqnarray}

\noindent If $S$, $V$ and $U$ are radial functions, the Dirac spinor is considered as
\begin{equation}\label{psi}
    \Psi_{km}(\vec{r})=\left(
                         \begin{array}{c}
                           \Psi_{+} \\
                           \Psi_{-} \\
                         \end{array}
                       \right)=
    \left(
      \begin{array}{c}
        \frac{if_{k}(r)}{r}\,Y_{km}(\hat{r}) \\
        \frac{g_{k}(r)}{r}\,Y_{-km}(\hat{r}) \\
      \end{array}
    \right)
\end{equation}

\noindent where $f_{k}$ and $g_{k}$ are the radial wave functions of the upper and lower components, res\-pec\-ti\-vely. $Y_{km}$ are the so-called spinor spherical harmonics. Here $k$ is the quantum number of the total angular momentum $j$ and it is related to the orbital momentum $l$ by $k=-(l+1)=-(j+1/2)$ for $j=l+1/2$ and $k=l=+(j+1/2)$ for $j=l-1/2$.

As shown in Ref. \cite{castro3}, using the following property $\vec{\sigma}\cdot\hat{r}Y_{km}=-Y_{-km}$, the equations (\ref{eqd3a}) and (\ref{eqd3b}) can be reduced to two coupled first order ordinary differential equations for the radial upper ($f_{k}$) and lower ($g_{k}$) components
\begin{eqnarray}
  \left[ \frac{d}{dr}+\frac{k}{r}-U(r) \right]f_{k}(r) &=& \left[ E+m-\Delta \right]g_{k}(r)\label{eqm1} \\
  \left[ \frac{d}{dr}-\frac{k}{r}+U(r) \right]g_{k}(r) &=& -\left[ E-m-\Sigma \right]f_{k}(r)\label{eqm2}
\end{eqnarray}

Under the discrete chiral transformations the spinor (\ref{psi}) becomes
\begin{equation}\label{psichi}
    \Psi_{\chi}=\gamma^{5}\Psi=\left(
                  \begin{array}{c}
                    \Psi_{-} \\
                    \Psi_{+} \\
                  \end{array}
                \right)
\end{equation}

\noindent This last result means that $\gamma^{5}$ interchanges the upper and lower components, thus $f$ turns into $-ig$, $g$ turns into $if$ and $k$ turns into $-k$.

\subsection{Spin and pseudospin symmetry}

Using the expression for $g_{k}$ obtained from (\ref{eqm1}) with $\Delta=0$ and $E\neq-m$ and inserting it in (\ref{eqm2}) we obtain
\begin{equation}\label{eq_spin}
    \left[ \frac{d^{2}}{dr^{2}}-\frac{k(k+1)}{r^{2}}+2k\frac{U(r)}{r}-U^{\prime}(r)-U^{2}(r) \right]f_{k}(r)=-\left( E-m-\Sigma \right)\left(E+m  \right)f_{k}(r)
\end{equation}

\noindent In a similar way, using the expression for $f_{k}$ obtained from (\ref{eqm2}) with $\Sigma=0$ and $E\neq m$ and inserting it in (\ref{eqm1}) we obtain
\begin{equation}\label{eq_pspin}
    \left[ \frac{d^{2}}{dr^{2}}-\frac{k(k-1)}{r^{2}}+2k\frac{U(r)}{r}+U^{\prime}(r)-U^{2}(r) \right]g_{k}(r)=-\left( E+m-\Delta \right)\left(E-m  \right)g_{k}(r)
\end{equation}

\noindent Therefore, either for $\Delta=0$ with $E\neq-m$ or $\Sigma=0$ with $E\neq m$ the solutions for this problem can be found by solving a Schr\"{o}dinger-like problem.

As discussed in the previous section, the discrete chiral transformation performs the changes $\Delta\rightarrow\Sigma$, $\Sigma\rightarrow\Delta$, $m\rightarrow-m$, $U\rightarrow-U$, $f\rightarrow-ig$, $g\rightarrow i f$ and $k\rightarrow-k$. At this stage, note that applying these changes in the equation (\ref{eqm1}) (or (\ref{eq_spin})) we obtain (\ref{eqm2}) (or (\ref{eq_pspin})). This means that we can take advantage of this kind of transformation and can obtain the solutions for $\Sigma=0$ from the $\Delta=0$ case. For instance, we can focus the discussion to this case $\Delta=0$, $\Sigma=c_{1}F_{1}(r)$ and $U=c_{2}F_{2}(r)$,  and the results for the case when $\Delta=c_{1}F_{1}(r)$, $\Sigma=0$ and $U=c_{2}F_{2}(r)$ can be obtained easily by just changing the sign of $m$, $c_{2}$ and $k$ in the relevant expressions.

\subsection{Dirac equation with Cornell potentials}

In the first instance, let us consider
\begin{equation}\label{simes}
    \Delta=0, \qquad \Sigma=a_{1}r+\frac{b_{1}}{r}, \qquad U=a_{2}r+\frac{b_{2}}{r}
\end{equation}

\noindent Substituting (\ref{simes}) into (\ref{eq_spin}) we get
\begin{equation}\label{effeq}
    \frac{d^{2}f_{k}(r)}{dr^{2}}+\left[ \mathcal{E}^{2}+\frac{a}{r}-br-cr^{2}-\frac{\lambda(\lambda+1)}{r^{2}} \right]f_{k}(r)=0
\end{equation}

\noindent where
\begin{equation}\label{eeff}
    \mathcal{E}^{2}=E^{2}-m^{2}+2a_{2}\left( k-b_{2}-\frac{1}{2} \right)
\end{equation}
\begin{equation}\label{a}
    a=-b_{1}\left( E+m \right)
\end{equation}
\begin{equation}\label{b}
    b=a_{1}\left( E+m \right)
\end{equation}
\begin{equation}\label{c}
    c=a_{2}^{2}
\end{equation}
\begin{equation}\label{lambda}
    \lambda=-\frac{1}{2}+\frac{1}{2}|2k+1-2b_{2}|
\end{equation}

\noindent The solution for (\ref{effeq}), with $c$ necessarily real and positive, is the solution of the Schr\"{o}dinger equation for the three-dimensional harmonic oscillator plus a Cornell potential. This novel potential is considered in \cite{leau} and \cite{heun}, but the authors misunderstood the full meaning of the potential and made a few erroneous calculations. We use this opportunity to present the correct solution to this problem in a more transparent way.

The solution close to the origin valid for all values of $\lambda$ can be written as being proportional to $r^{\lambda+1}$. Putting
\begin{equation}\label{ansatz}
f(r)=r^{\lambda+1}\exp \left( -\frac{\sqrt{c}}{2}\,r^{2}-\frac{b}{2\sqrt{c}}%
\,r\right) \phi(r)
\end{equation}

\noindent and introducing the following new variable and parameters%
\begin{equation}\label{newv}
x=\sqrt[4]{c}\,r\,,\quad \omega =2\lambda+1\,,\quad \rho =\frac{b}{\sqrt[4]{c^{3}}}%
\,,\quad \tau =\frac{b^{2}+4\,c\,\varepsilon ^{2}}{4\sqrt{c^{3}}}
\end{equation}%
one finds that the solution for all $r$ can be expressed as a solution of
the biconfluent Heun differential equation \cite{heun}.
\begin{equation}\label{heuneq}
x\,\frac{d^{2}\phi}{dx^{2}}+(\omega +1-\rho x-2x^{2})\,\frac{d\phi}{dx}+\left[
\left( \tau -\omega -2\right) x-\Theta \right] \phi=0
\end{equation}%

\noindent with%
\begin{equation}\label{Delta}
\Theta =\frac{1}{2}\left[ \delta +\rho \left( \omega +1\right) \right]
\end{equation}

\noindent where $\delta=-\frac{2a}{\sqrt[4]{c}}$. The expression for $\omega$, $\tau$ and $\delta$ are very different from that given in \cite{leau} and \cite{heun}. The reason for this disagreement are mistakes in those references. The biconfluent Heun differential
equation has a regular singularity at $x=0$ and an irregular singularity at $%
x=\infty $. The solution regular at the origin is given by%
\begin{equation}\label{sol}
N\left( \omega ,\rho ,\tau ,\delta ;x\right) =\sum_{j=0}^{\infty }\frac{%
\Gamma \left( \omega+1\right) }{\Gamma \left( \omega +1+j\right) }\,\frac{A_{j}}{j!}%
\,x^{j}
\end{equation}%
where $\Gamma \left( z\right) $ is the gamma function, $A_{0}=1$, $A_{1}=\Theta$ and the
remaining coefficients of the the series expansion, for $\rho \neq 0$,
satisfy the three-term recurrence relation:
\begin{equation}\label{rec}
A_{j+2}=\left[ \left( j+1\right) \rho +\Theta \right] A_{j+1}-\left(
j+1\right) \left( j+\omega +1\right) \left( \tau -\omega -2-2j\right) A_{j}
\end{equation}%
The series is convergent for $x$ in the range $[0,\infty )$ and tends to $%
e^{x^{2}}$ as $x\rightarrow \infty $. In fact, $\phi$
presents polynomial solutions of degree $n$ when $\tau=\omega +2+2n$ and $A_{n+1}=0$.
Therefore, using the condition $\tau=\omega +2+2n$ we obtain
\begin{equation}\label{exe}
    \mathcal{E}^{2}=(2n+2\lambda+3)|a_{2}|-\frac{b^{2}}{4a_{2}^{2}}
\end{equation}

\noindent Substituting (\ref{eeff}) and (\ref{b}) into (\ref{exe}), we obtain the spectrum for $\Delta=0$
\begin{equation}\label{ener}
    E=-\frac{m_{E}}{a_{E}}\pm\frac{m_{E}}{a_{E}}\sqrt{1-\frac{a_{E}}{m_{E}^{2}}\left[ 2a_{2}\left( k-b_{2}-\frac{1}{2} \right)-
    \left( 2n+2\lambda+3 \right)|a_{2}|+\left( a_{E}-2 \right)m^{2} \right]}
\end{equation}

\noindent where $m_{E}=\frac{ma_{1}^2}{4a_{2}^{2}}$ and $a_{E}=1+\frac{a_{1}^{2}}{4a_{2}^{2}}$. Note that, at first view (\ref{ener}) is independent of the value of $b_{1}$. Now we focus attention on the condition $A_{n+1}=0$, this condition provides a constraint on the value of $b_{1}$. For instance, $n=0$ implies that $A_{1}=\Theta=0$. In this specific case, we obtain
\begin{equation}\label{vinb1}
    b_{1}=-\frac{a_{1}}{2|a_{2}|}\,\left( 1+|1-2b_{2}+2k| \right)
\end{equation}

\noindent the constraint (\ref{vinb1}) involving specific values of $a_{1}$, $a_{2}$, $b_{2}$ and the quantum number of the total angular momentum. At this stage, we can see a peculiar behavior of the parameter $b_{1}$. Initially, $b_{1}$ is arbitrary but during the procedure to obtain the quantization condition the value of $b_{1}$ is restricted by (\ref{vinb1}). This last result implies that the parameter $b_{1}$ in (\ref{simes}) should satisfy the constrain (\ref{vinb1}) in order to obtain the quantization condition. Therefore, we can conclude that the spectrum (\ref{ener}) depends implicity of $b_{1}$, due to there is a link between the parameters $a_{1}$, $a_{2}$, $b_{1}$, $b_{2}$ and $k$.

Now, let us consider the case
\begin{equation}\label{simeps}
    \Delta=a_{1}r+\frac{b_{1}}{r}, \qquad \Sigma=0, \qquad U=a_{2}r+\frac{b_{2}}{r}
\end{equation}

\noindent As referred before, we can take advantage of the discrete chiral transformation. We recall that this case can be obtained easily by just the changes $m\rightarrow-m$, $U(r)\rightarrow-U(r)$ and $k\rightarrow-k$ in the relevant expressions. We can see that the change in $U(r)$ implies that $a_{2}\rightarrow-a_{2}$ and $b_{2}\rightarrow-b_{2}$. Therefore, the spectrum for $\Sigma=0$ is given by
\begin{equation}\label{ener2}
    E=\frac{m_{E}}{a_{E}}\pm\frac{m_{E}}{a_{E}}\sqrt{1-\frac{a_{E}}{m_{E}^{2}}\left[ 2a_{2}\left( k-b_{2}+\frac{1}{2} \right)-
    \left( 2n+2\lambda+3 \right)|a_{2}|+\left( a_{E}-2 \right)m^{2} \right]}
\end{equation}

\noindent where $\lambda=-\frac{1}{2}+\frac{1}{2}|1-2k+2b_{2}|$.

\section{conclusions}

We have addressed the behavior of the Dirac equation with scalar ($S$), vector ($V$) and tensor ($U$) interactions under the $\gamma^{5}$ discrete chiral transformation. We showed that is possible obtain from a simple way solutions of the Dirac equation for $\Sigma=0$ (pseudospin symmetry) from the $\Delta=0$ (spin symmetry) case using symmetry arguments. As a application, we have considered scalar, vector and tensor Cornell radial potentials. For this case, the radial equation is mapped into a Scr\"{o}dinger-like equation embedded in a three-dimensional harmonic oscillator plus a Cornell potential. We found the correct solution for this problem. Our results are definitely useful because shed some light in a issue that has not been reported in the literature. Additionally, the correct solution for the Cornell potential may be useful due to a wide application in several physical problems.

\begin{acknowledgments}
 L.B. Castro would like to thank the referee for useful comments and suggestions. This work was supported by means of funds provided by CAPES.
\end{acknowledgments}

\end{document}